\documentclass[twocolumn]{aastex631}

\usepackage{amsmath}
\usepackage{bm}
\usepackage{multirow}

\graphicspath{{./}{Figures/}}
\shorttitle{Massive neutron stars with small radii}
\shortauthors{Miyatsu, Cheoun, Kim, and Saito}

\begin{document}

% \title{Massive neutron stars with small radii in relativistic mean-field models with isoscalar- and isovector-meson mixing}
\title{Massive neutron stars with small radii in relativistic mean-field models optimized to nuclear ground states}

\correspondingauthor{Tsuyoshi Miyatsu}
\email{tsuyoshi.miyatsu@ssu.ac.kr}

\author[0000-0001-9186-8793]{Tsuyoshi Miyatsu}
\affiliation{Department of Physics and OMEG Institute, Soongsil University, Seoul 06978, Republic of Korea}

\author[0000-0001-7810-5134]{Myung-Ki Cheoun}
\affiliation{Department of Physics and OMEG Institute, Soongsil University, Seoul 06978, Republic of Korea}

\author{Kyungsik Kim}
\affiliation{School of Liberal Arts and Sciences, Korea Aerospace University, Goyang 10540, Republic of Korea}

\author[0000-0002-8563-9262]{Koichi Saito}
\affiliation{Department of Physics, Faculty of Science and Technology, Tokyo University of Science, Noda 278-8510, Japan}

\begin{abstract}
  We present an equation of state (EoS) for neutron stars using the relativistic mean-field model with isoscalar- and isovector-meson mixing.
  Taking into account the results of the neutron skin thickness, $R_{\rm skin}$, of $^{208}$Pb reported by the PREX collaboration, the dimensionless tidal deformability of a canonical neutron star observed from GW170817, and a $2.6$ $M_{\odot}$ compact star implied by the secondary component of GW190814, a new effective interaction is constructed so as to reproduce the saturation condition of nuclear matter and the ground-state properties of finite, closed-shell nuclei.
  We find that the neutron star EoS exhibits the rapid stiffening around twice the nuclear saturation density, which is caused by the soft nuclear symmetry energy, $E_{\rm sym}$.
  It is also noticeable that the thick $R_{\rm skin}$ from the PREX-2 experiment can be achieved with the small slope parameter of $E_{\rm sym}$ stemming from the isoscalar-meson mixing.
  Thus, we speculate that the secondary component of GW190814 is the heaviest neutron star ever discovered.
\end{abstract}


\section{Introduction} \label{sec:introduction}

The astrophysical phenomena concerning compact stars as well as the properties of finite nuclei and nuclear matter are determined by the nuclear equation of state (EoS), characterized by the relation between the energy density and pressure of the system.
Owing to the precise observations of neutron stars, such as the Shapiro delay measurement of a binary millisecond pulsar J1614$-$2230~\citep{Demorest:2010bx,NANOGrav:2017wvv} and the radius measurement of PSR J0740$+$6620 from Neutron Star Interior Composition Explorer (NICER) and from X-ray Multi-Mirror (XMM-Newton) Data~\citep{Miller:2021qha}, theoretical studies have been currently performed more than ever to elucidate neutron star physics through the EoS for dense matter.

In addition, the direct detection of gravitational-wave (GW) signals from a binary neutron star merger, GW170817, observed by Advanced LIGO and Advanced Virgo detectors, have placed stringent restrictions on the mass--radius relation of neutron stars~\citep{LIGOScientific:2017vwq,LIGOScientific:2018cki,LIGOScientific:2018hze}.
Especially, the tidal deformability of a neutron star~\citep{Hinderer:2007mb,Hinderer:2009ca} plays an important role in constructing the EoS for neutron star matter~\citep{Annala:2017llu,Lim:2018bkq,Raithel:2018ncd}.
Moreover, the secondary component of GW190814 with the mass of $2.6$ $M_{\odot}$ poses another fascinating question whether it is the lightest black hole or the heaviest neutron star~\citep{LIGOScientific:2020zkf}.
Recently, several ideas on this new topic have been proposed in astrophysics.
\citet{Fattoyev:2020cws} have insisted that the $2.6$ $M_{\odot}$ object is likely to be the lightest black hole ever discovered using the nucleonic EoS.
In contrast, others support the possibility of the secondary object of GW190814 as a neutron star~\citep{Huang:2020cab,Biswas:2020xna,Bombaci:2020vgw,Dexheimer:2020rlp,Drischler:2020fvz,Ferreira:2021pni,Lim:2020zvx,Lopes:2021yga,Miao:2021nuq,Wu:2021rxw,Wang:2022ovh}.

On the other hand, a critical issue has been %discovered
raised in nuclear physics since the accurate determination of neutron skin thickness, $R_{\rm skin}$, of $^{208}$Pb through parity-violation in electron scattering by the PREX collaboration~\citep{PREX:2021umo}.
It is a very challenging task to understand the result from the PREX-2 experiment because the measured $R_{\rm skin}$ is remarkably larger than those expected theoretically.
The calculations of $R_{\rm skin}$ are generally settled by the nuclear symmetry energy, $E_{\rm sym}$, and its slope parameter, $L$, near the nuclear saturation density, $\rho_{0}$, mainly characterizing the properties of isospin-asymmetric nuclear matter and heavy nuclei.
Many active discussions on $L$ via the nuclear EoS have been performed since the PREX-2 experiment.
Some calculations favor the small value of $L$ less than around 70 MeV, for example, the energy density functionals or the Bayesian inference~\citep{Drischler:2020hwi,Xu:2020fdc,Biswas:2021yge,Essick:2021ezp,Essick:2021kjb,Huth:2020ozf,Reinhard:2021utv}, whereas the others employ the larger $L$ than 100 MeV~\citep{SRIT:2021gcy,Reed:2021nqk} to account for the PREX-2 data, the experimental analyses of heavy-ion collisions, and the astrophysical observations of neutron stars.
However, it is still hard to determine the exact value of $L$~\citep{Piekarewicz:2021jte}.

In our previous study~\citep{Miyatsu:2022wuy}, we have developed the relativistic mean-field (RMF) model with nonlinear couplings by introducing the isoscalar- and isovector-meson mixing, $\sigma^{2}\bm{\delta}^{2}$ and $\omega_{\mu}\omega^{\mu}\bm{\rho}_{\nu}\bm{\rho}^{\nu}$.
It has been found that the quartic interaction due to the scalar mesons has a large influence on the radius and tidal deformability of a neutron star.
In particular, the $\sigma$--$\delta$ mixing could simultaneously reproduce the dimensionless tidal deformability of a canonical $1.4$ $M_{\odot}$ neutron star, $\Lambda_{1.4}$, measured by the GW signals from GW170817 and GW190814.
The effects of hyperons, deconfinement, and the speed of sound in the core of a neutron star have been also studied using the same RMF model in SU(6) spin-flavor symmetry~\citep{Aguirre:2022zxn}.
We note that \citet{Li:2022okx} have also explored the possibility of the thick $R_{\rm skin}$ of $^{208}$Pb from the PREX-2 experiment by the inclusion of the $\delta$ meson and its mixing.

In the present study, we construct a new effective interaction based on the RMF model with isoscalar- and isovector-meson mixing to explain both data from stable nuclear ground states and astrophysical observations of neutron stars.
In particular, we focus on the possibility of massive neutron stars with small radii in order to account for the thick $R_{\rm skin}$ of $^{208}$Pb measured by the PREX-2 experiment, the small $\Lambda_{1.4}$ observed from GW170817, and a hypermassive neutron star with the mass of $2.6$ $M_{\odot}$ implied by the secondary component of GW190814.

This paper is organized as follows.
A brief summary of the RMF model with nonlinear couplings is provided in Section~\ref{sec:formalism}.
Numerical results and detailed discussions concerning features of the new effective interaction are presented in Section~\ref{sec:results}.
Finally, we give a summary in Section~\ref{sec:summary}.

\section{Formalism} \label{sec:formalism}

The recently updated Lagrangian density in RMF approximation is employed to construct the EoS for nuclear and neutron star matter~\citep{Miyatsu:2022wuy}.
We introduce the isoscalar ($\sigma$ and $\omega^{\mu}$) and isovector ($\bm{\delta}$ and $\bm{\rho}^{\mu}$) mesons as well as nucleons ($N=p,n$).
The interacting Lagrangian density is then given by
\begin{align}
  \mathcal{L}_{\rm int}
  & = \sum_{N}\bar{\psi}_{N}\bigl[g_{\sigma}\sigma
    - g_{\omega}\gamma_{\mu}\omega^{\mu}
    + g_{\delta}\bm{\delta}\cdot\bm{\tau}_{N}
    \nonumber \\
  & - g_{\rho}\gamma_{\mu}\bm{\rho}^{\mu}\cdot\bm{\tau}_{N}\bigr]\psi_{N}
    - U_{\rm NL}(\sigma,\omega^{\mu},\bm{\delta},\bm{\rho}^{\nu}),
    \label{Lint}
\end{align}
where $\psi_{N}$ is the nucleon field and $\bm{\tau}_{N}$ is its isospin matrix.
The meson-nucleon coupling constants are respectively denoted by $g_{\sigma}$, $g_{\omega}$, $g_{\delta}$, and $g_{\rho}$.
A nonlinear potential is here supplemented as
\begin{align}
  U_{\rm NL}(\sigma,\omega^{\mu},\bm{\delta},\bm{\rho}^{\nu})
  & = \frac{1}{3}g_{2}\sigma^{3} + \frac{1}{4}g_{3}\sigma^{4} - \frac{1}{4}c_{3}\left(\omega_{\mu}\omega^{\mu}\right)^{2}
    \nonumber \\
  & - \Lambda_{\sigma\delta}\sigma^{2}\bm{\delta}^{2}
    - \Lambda_{\omega\rho}\left(\omega_{\mu}\omega^{\mu}\right)\left(\bm{\rho}_{\nu}\cdot\bm{\rho}^{\nu}\right),
    \label{eq:NLpot}
\end{align}
with five coupling constants and mixing parameters, $g_{2}$, $g_{3}$, $c_{3}$, $\Lambda_{\sigma\delta}$, and $\Lambda_{\omega\rho}$~\citep{Boguta:1977xi,Todd-Rutel:2005yzo,Miyatsu:2013yta,Miyatsu:2022wuy,Zabari:2018tjk}.
For describing the characteristics of finite nuclei, the Coulomb interaction, $\mathcal{L}_{C}=-e\bar{\psi}_{p}\gamma_{\mu}A^{\mu}\psi_{p}$, for photon $A^{\mu}$ is also taken into consideration.

\section{Results and discussions} \label{sec:results}

%%%%%%%%%%%%%%%%%%%%%%%%%%%%%%%%%%%%%%%%%%%%%%%%%%%%%%%%%%%%%%%%%%%%%%%%%%%%%%%% 
\begin{table}[t!]
  \caption{\label{tab:parameters}
    Model parameters and properties of nuclear matter at $\rho_{0}$.}
  \begin{ruledtabular}
    \begin{tabular}{lrlr}
      \multicolumn{2}{c}{Model parameters}  & \multicolumn{2}{c}{Bulk properties} \\
      \colrule
      $g_{\sigma}$             &      9.45  & $\rho_{0}$ (fm$^{-3}$) &      0.15  \\
      $g_{\omega}$             &     11.95  & $M_{N}^{\ast}/M_{N}$   &      0.64  \\
      $g_{\delta}$             &      6.14  & $E_{0}$ (MeV)          &   $-16.45$ \\
      $g_{\rho}$               &      7.19  & $K_{0}$ (MeV)          &    280.00  \\
      $g_{2}$ (fm$^{-1}$)      &      9.98  & $J_{0}$ (MeV)          &   $-66.98$ \\
      $g_{3}$                  &   $-21.47$ & $E_{\rm sym}$ (MeV)    &     34.55  \\
      $c_{3}$                  &      0.00  & $L$ (MeV)              &     50.00  \\
      $\Lambda_{\sigma\delta}$ &     87.00  & $K_{\rm sym}$ (MeV)    &  $-384.43$ \\
      $\Lambda_{\omega\rho}$   &    102.61  & $J_{\rm sym}$ (MeV)    &  $-533.43$ \\
    \end{tabular}
  \end{ruledtabular}
  \tablecomments{The nucleon and meson masses are respectively fixed by $M_{N}=939.00$ MeV, $m_{\sigma}=496.50$ MeV, $m_{\omega}=782.66$ MeV, $m_{\delta}=980.00$ MeV, and $m_{\rho}=775.26$ MeV~\citep{ParticleDataGroup:2020ssz}.
    The bulk properties of nuclear matter are given by the coefficients based on the expansion of isospin-asymmetric nuclear EoS with power series in the isospin asymmetry around $\rho_{0}$~\citep{Stone:2014wza,Choi:2020eun}.}
\end{table}
%%%%%%%%%%%%%%%%%%%%%%%%%%%%%%%%%%%%%%%%%%%%%%%%%%%%%%%%%%%%%%%%%%%%%%%%%%%%%%%% 
A new effective interaction can reproduce the saturation properties of nuclear matter and the characteristics of finite nuclei.
The resulting parameter set---henceforth referred to as the OMEG (Origin of Matter and Evolution of Galaxies) interaction---is listed in Table~\ref{tab:parameters}.
There are three modifications from the previous work~\citep{Miyatsu:2022wuy}.
We change $\rho_{0}$ into 0.15 fm$^{-3}$ in order to describe the weak charge and baryon density profiles of $^{208}$Pb as well as its experimental charge density profile, based on the PREX-2 result~\citep{Horowitz:2020evx,PREX:2021umo}.
In addition, because of the strong correlation between $\Lambda_{1.4}$ and $M_{N}^{\ast}$~\citep{Hornick:2018kfi,Choi:2020eun}, the effective mass ratio of nucleon at $\rho_{0}$ is fixed by $M_{N}^{\ast}/M_{N}=0.64$ to accomplish the observed $\Lambda_{1.4}$ from GW170817, which is within the mass range in \citet{Li:2018lpy}.
Furthermore, the $\delta$-$N$ coupling constant is tuned so as to enhance the $\delta$-meson effect on the isospin-asymmetric nuclear EoS.
In the present study, we set $g_{\delta}^{2}/4\pi=3$, at which the curvature parameter of nuclear symmetry energy, $K_{\rm sym}$, attains a minimum limit~\citep{Miyatsu:2022wuy}.
Note that the $\sigma$-$\delta$ mixing parameter, $\Lambda_{\sigma\delta}=87$, is adopted to ensure the matter stability of charge fluctuations.
When $\Lambda_{\sigma\delta}$ is larger than the current value, then nuclear matter becomes unstable at high densities and the phase transition should be considered~\citep{Kubis:2020ysv}.

%%%%%%%%%%%%%%%%%%%%%%%%%%%%%%%%%%%%%%%%%%%%%%%%%%%%%%%%%%%%%%%%%%%%%%%%%%%%%%%% 
\begin{table}[t!]
  \caption{\label{tab:finite}
    Theoretical predictions for ground-state properties of several closed-shell nuclei.}
  \begin{ruledtabular}
    \begin{tabular}{lccccr}
      \multirow{2}{*}{Nucleus} & \multicolumn{2}{c}{$B/A$ (MeV)} & \multicolumn{2}{c}{$R_{\rm ch}$ (fm)} & \multirow{2}{*}{$R_{\rm skin}$ (fm)} \\
      \cline{2-3}\cline{4-5}
      \          & Theory & Exp. & Theory & Exp.    &       \ \\
      \colrule
      $^{16}$O   &   8.03 & 7.98 &   2.75 &    2.70 & $-0.03$ \\
      $^{40}$Ca  &   8.58 & 8.55 &   3.48 &    3.48 & $-0.05$ \\
      $^{48}$Ca  &   8.59 & 8.67 &   3.50 &    3.48 &   0.20  \\
      $^{68}$Ni  &   8.68 & 8.68 &   3.89 &    3.89 &   0.22  \\
      $^{90}$Zr  &   8.71 & 8.71 &   4.28 &    4.27 &   0.09  \\
      $^{100}$Sn &   8.25 & 8.25 &   4.49 & \nodata & $-0.08$ \\
      $^{132}$Sn &   8.34 & 8.35 &   4.72 &    4.71 &   0.29  \\
      $^{208}$Pb &   7.90 & 7.87 &   5.51 &    5.50 &   0.23  \\
    \end{tabular}
  \end{ruledtabular}
  \tablecomments{Experimental data for the binding energy per nucleon, $B/A$, and charge radius, $R_{\rm ch}$, are referred to \citet{Wang:2021xhn} and \citet{Angeli:2013epw}, respectively.}
\end{table}
%%%%%%%%%%%%%%%%%%%%%%%%%%%%%%%%%%%%%%%%%%%%%%%%%%%%%%%%%%%%%%%%%%%%%%%%%%%%%%%% 
The binding energies and charge radii of several closed-shell nuclei for the OMEG interaction are summarized in Table~\ref{tab:finite}.
Although the small value of $L$ generally gives thin $R_{\rm skin}$ in the RMF models, the $\delta$ meson and its mixing enable us to obtain relatively thick $R_{\rm skin}$ even in the calculations with small $L$.
We here find that the OMEG interaction provides $R_{\rm skin}=0.23$ fm with $L=50$ MeV, which satisfies the PREX-2 data of $0.283\pm0.071$ fm~\citep{PREX:2021umo}.

%%%%%%%%%%%%%%%%%%%%%%%%%%%%%%%%%%%%%%%%%%%%%%%%%%%%%%%%%%%%%%%%%%%%%%%%%%%%%%%% 
\begin{figure}[t!]
  \plotone{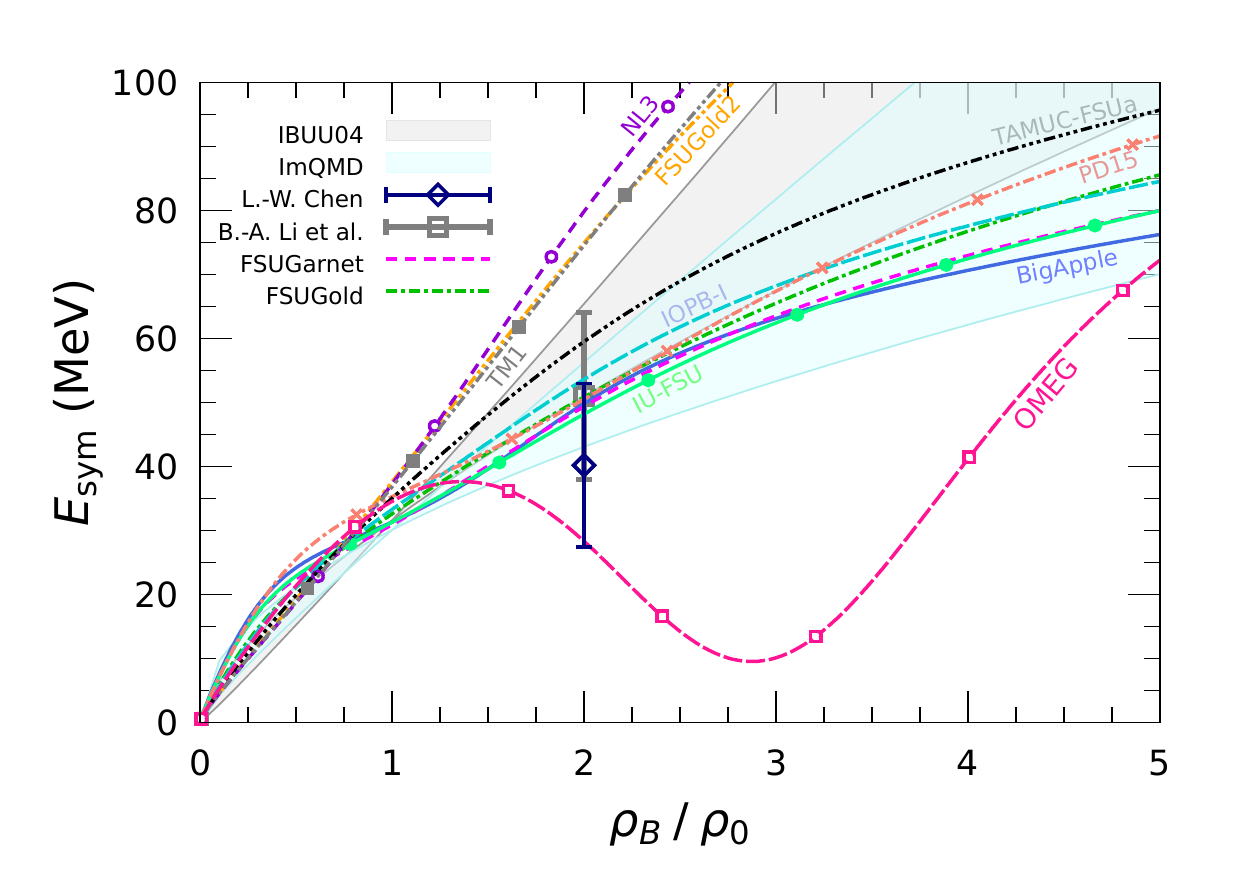}
  \caption{\label{fig:Esym}
    Nuclear symmetry energy, $E_{\rm sym}$, as a function of the baryon density ratio, $\rho_{B}/\rho_{0}$.
    The constraints from analyses of heavy-ion collision data using the isospin-dependent Boltzmann--Uehling--Uhlenbec (IBUU04) and improved quantum molecular dynamics (ImQMD) transport models are presented~\citep{Chen:2004si,Li:2005jy,Tsang:2008fd}.
    We also show the constraints on the magnitude of $E_{\rm sym}$ at $2\rho_{0}$: $E_{\rm sym}(2\rho_{0})\simeq40.2\pm12.8$ MeV based on microscopic calculations with various energy density functionals~\citep{Chen:2015gba}, and $E_{\rm sym}(2\rho_{0})\simeq51\pm13$ MeV from nine new analyses of neutron star observables since GW170817~\citep{Li:2021thg}.}
\end{figure}
%%%%%%%%%%%%%%%%%%%%%%%%%%%%%%%%%%%%%%%%%%%%%%%%%%%%%%%%%%%%%%%%%%%%%%%%%%%%%%%% 
The density dependence of nuclear symmetry energy, $E_{\rm sym}$, for the OMEG interaction is displayed in Figure~\ref{fig:Esym}.
Various theoretical calculations using the well-calibrated parameter sets based on the RMF models are also presented: BigApple~\citep{Fattoyev:2020cws}, FSUGarnet~\citep{Chen:2014mza}, FSUGold~\citep{Fattoyev:2010tb}, FSUGold2~\citep{Chen:2014sca}, IOPB-I~\citep{Kumar:2017wqp}, IU-FSU~\citep{Fattoyev:2010mx}, NL3~\citep{Lalazissis:1996rd}, PD15~\citep{Liliani:2021jne}, TAMUC-FSUa~\citep{Fattoyev:2013yaa,Piekarewicz:2013bea}, and TM1~\citep{Sugahara:1993wz}.

As explained in the previous study~\citep{Miyatsu:2022wuy}, $E_{\rm sym}$ for the OMEG interaction shows an inflection point above $\rho_{0}$, the rapid reduction around $1.5\rho_{0}$--$2.8\rho_{0}$, and the suppression above $3\rho_{0}$, which are caused by the strong $\sigma$--$\delta$ mixing.
This behavior is similar to the cusp in $E_{\rm sym}$ using the skyrmion crystal approach~\citep{Ma:2021nuf,Lee:2021hrw} and to the results in the Skyrme Hartree-Fock calculations~\citep{Chen:2005ti}.

%%%%%%%%%%%%%%%%%%%%%%%%%%%%%%%%%%%%%%%%%%%%%%%%%%%%%%%%%%%%%%%%%%%%%%%%%%%%%%%% 
\begin{figure}[t!]
  \plotone{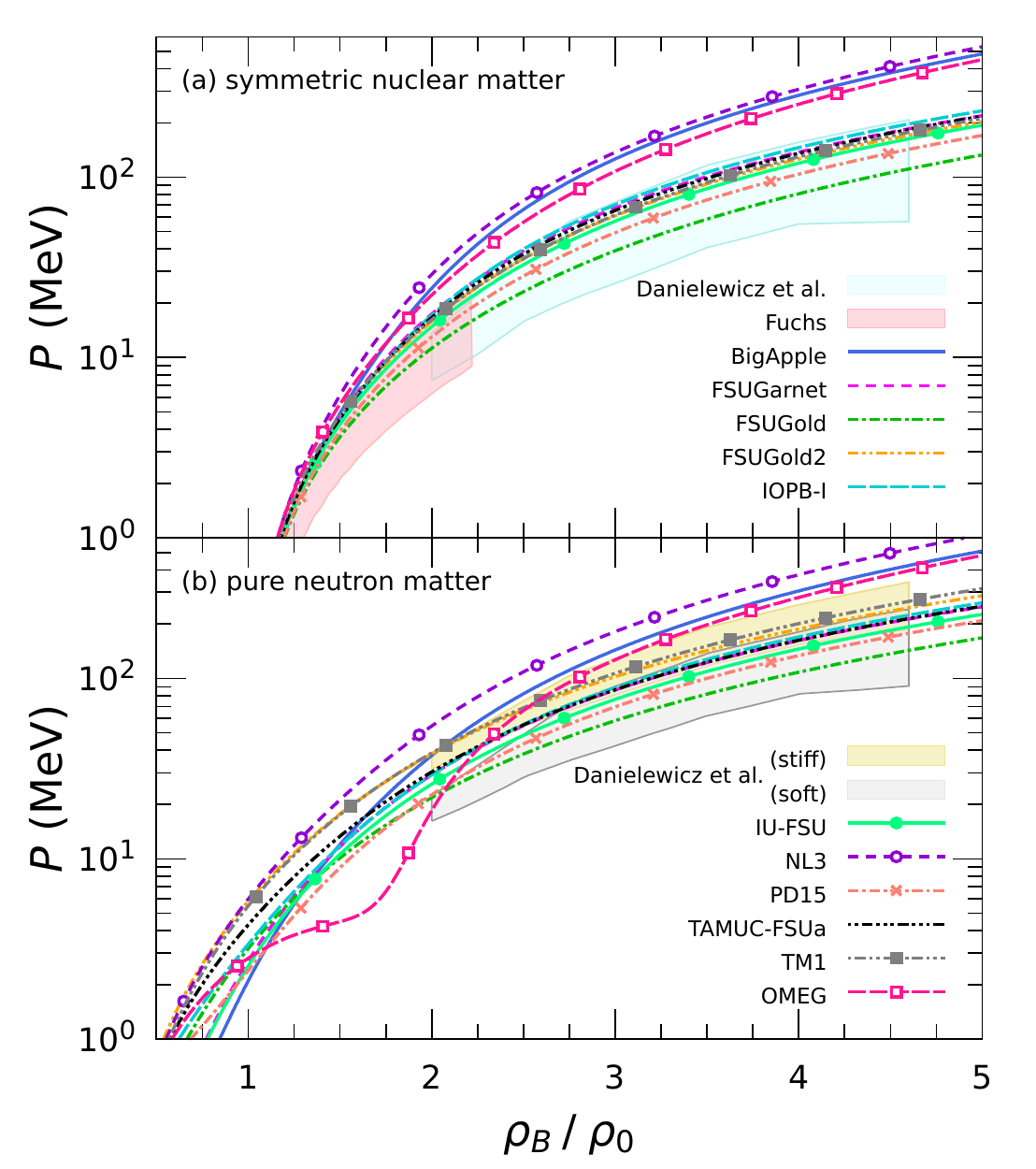}
  \caption{\label{fig:Pressure}
    EoS---Pressure, $P$, as a function of $\rho_{B}/\rho_{0}$---for (a) symmetric nuclear matter and (b) pure neutron matter.
    The shaded areas represent the constraints from elliptical flow data~\citep{Danielewicz:2002pu} and kaon production data~\citep{Fuchs:2005zg}.}
\end{figure}
%%%%%%%%%%%%%%%%%%%%%%%%%%%%%%%%%%%%%%%%%%%%%%%%%%%%%%%%%%%%%%%%%%%%%%%%%%%%%%%% 
The density dependence of pressure, $P$, is depicted in Figure~\ref{fig:Pressure}.
The EoS for the OMEG interaction is rather stiff in symmetric nuclear matter, similar to those for the BigApple and NL3 interactions.
In contrast, in pure neutron matter, the OMEG interaction shows a slow growth above $\rho_{0}$ and then a sharp increase around $2\rho_{0}$.
At high densities, it almost satisfies the constraint from elliptical flow data~\citep{Danielewicz:2002pu}.
Here, we can verify that the phase transition due to the matter instability does not occur even at high densities because the condition of $dP/d\rho_{B}>0$ is ensured in pure neutron matter.

Since the discovery of PSR J1614$-$2230 with the mass of $1.908\pm0.016$ $M_{\odot}$~\citep{Demorest:2010bx,NANOGrav:2017wvv}, the EoS for neutron stars has been constructed so as to support $2.0$ $M_{\odot}$.
However, PSR J0952$-$0607, which has the largest well-measured mass of $2.35\pm0.17$ $M_{\odot}$ with small modeling uncertainties, has been reported very recently~\citep{Romani:2022jhd}.
Furthermore, we have no reason to ignore the possibility of the secondary object of GW190814 as a neutron star~\citep{LIGOScientific:2020zkf}.
Therefore, at present, the maximum mass of a neutron star, $M_{\rm max}$, namely the Tolman--Oppenheimer--Volkoff (TOV) limit, might be larger than ever.

%%%%%%%%%%%%%%%%%%%%%%%%%%%%%%%%%%%%%%%%%%%%%%%%%%%%%%%%%%%%%%%%%%%%%%%%%%%%%%%% 
\begin{figure}[t!]
  \plotone{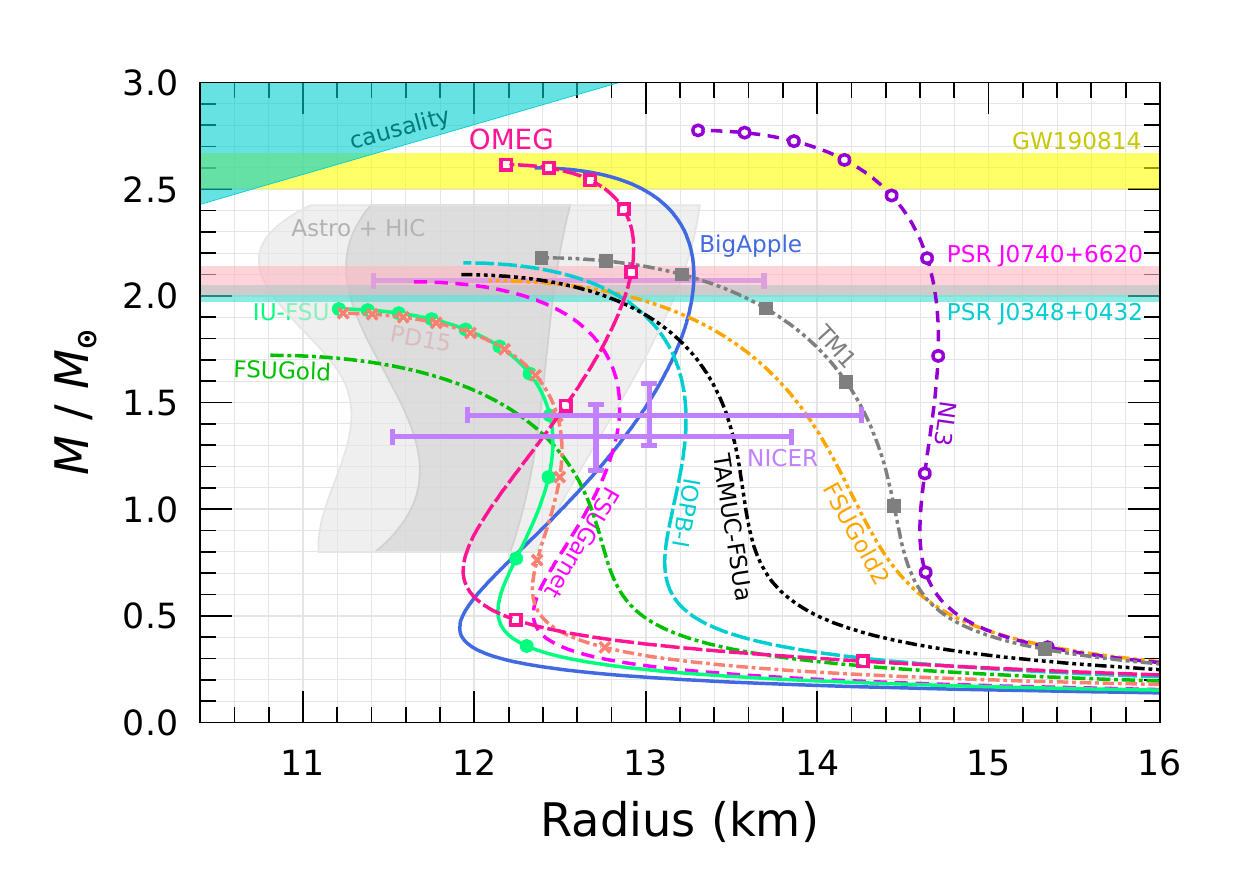}
  \caption{\label{fig:MR}
    Mass--radius relations of neutron stars.
    The observational data are supplemented by the constraints from PSR J0030$+$0451 by a NICER view ($1.44_{-0.14}^{+0.15}$ ${M}_{\odot}$ and $13.02_{-1.06}^{+1.24}$ km~\citep{Miller:2019cac} and $1.34_{-0.16}^{+0.15}$ ${M}_{\odot}$ and $12.71_{-1.19}^{+1.14}$ km~\citep{Riley:2019yda}), PSR J0348$+$0432 ($2.01\pm0.04$ $M_{\odot}$)~\citep{Antoniadis:2013pzd}, PSR J0740$+$6620 ($2.072^{+0.067}_{-0.066}$ $M_{\odot}$ and $12.39^{+1.30}_{-0.98}$ km)~\citep{NANOGrav:2019jur,Fonseca:2021wxt,Riley:2021pdl}, and the secondary object of GW190814 ($2.59^{+0.08}_{-0.09}$ $M_{\odot}$)~\citep{LIGOScientific:2020zkf}.
    The recent theoretical restriction using Bayesian inference is also shown in the shaded region~\citep{Huth:2021bsp}.}
\end{figure}
%%%%%%%%%%%%%%%%%%%%%%%%%%%%%%%%%%%%%%%%%%%%%%%%%%%%%%%%%%%%%%%%%%%%%%%%%%%%%%%% 
The mass--radius relations of neutron stars are presented in Figure~\ref{fig:MR}.
We here employ the EoS for nonuniform matter in the crust region, where nuclei are taken into account using the Thomas--Fermi calculation~\citep{Miyatsu:2013hea,Miyatsu:2015kwa}.
It is found that only the BigApple and OMEG interactions account for both constraints: $M_{\rm max}\ge2.35$ $M_{\odot}$ and the neutron star radii from J0030$+$0451~\citep{Miller:2019cac,Riley:2019yda} and from PSR J0740$+$6620~\citep{Riley:2021pdl}.
We here argue that the TOV limit can reach $2.6$ $M_{\odot}$, and thus the secondary component of GW190814 might be the heaviest neutron star.

We notice that the nonlinear $\omega$ self-coupling in Equation~\eqref{eq:NLpot} seems to be no longer necessary for supporting a hypermassive neutron star because its coupling constant, $c_{3}$, is very small and zero for the BigApple and OMEG interactions, respectively.
Moreover, the two interactions have a unique feature of $R_{2.0}>R_{1.4}$, where $R_{2.0(1.4)}$ is the neutron star radius with the mass of $2.0(1.4)$ $M_{\odot}$, as mentioned in \citet{Drischler:2021bup}.
Furthermore, the OMEG interaction is consistent with the theoretical restriction using Bayesian inference based on the combined data from astrophysical multi-messenger observations of neutron stars and from heavy-ion collisions of gold nuclei at relativistic energies with microscopic calculations~\citep{Huth:2021bsp}.

%%%%%%%%%%%%%%%%%%%%%%%%%%%%%%%%%%%%%%%%%%%%%%%%%%%%%%%%%%%%%%%%%%%%%%%%%%%%%%%% 
\begin{figure}[t!]
  \plotone{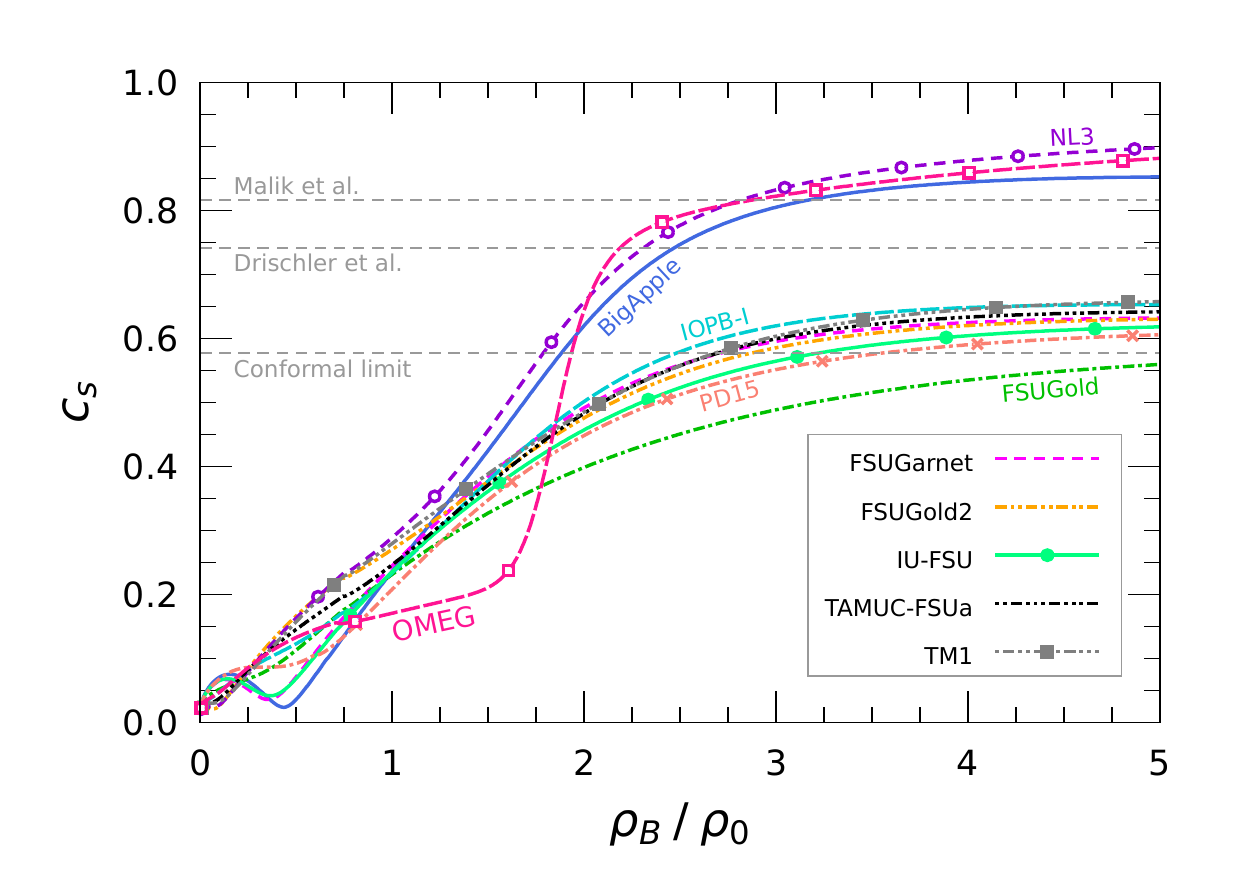}
  \caption{\label{fig:sound}
    Speed of sound dependence on $\rho_{B}/\rho_{0}$.
    The gray dashed lines show the theoretical indications: the conformal limit, $c_{s}=1/\sqrt{3}$, based on generic conditions for stable hybrid stars~\citep{Alford:2013aca} and the borders, $c_{s}=\sqrt{0.55}$ and $\sqrt{2/3}$, suggested by massive neutron stars~\citep{Drischler:2020fvz,Malik:2022zol}.}
\end{figure}
%%%%%%%%%%%%%%%%%%%%%%%%%%%%%%%%%%%%%%%%%%%%%%%%%%%%%%%%%%%%%%%%%%%%%%%%%%%%%%%% 
In Figure~\ref{fig:sound}, the speed of sound, $c_{s}=\sqrt{dP/d\varepsilon}$, in neutron star matter is illustrated as a function of $\rho_{B}/\rho_{0}$.
Because any exotic degrees of freedom in the core of a neutron star are not included, $c_{s}$ reaches a plateau at high densities.
At the central density of $M_{\rm max}$, the BigApple, NL3, and OMEG interactions exceed $c_{s}\simeq0.85$, while the others lie around the conformal limit~\citep{Alford:2013aca}.
It is thus found that $c_{s}$ is much larger than the conformal limit in the core of a hypermassive neutron star.
For the OMEG interaction, the rapid growth of $c_{s}$ occurs around $2\rho_{0}$, where the EoS for pure neutron matter changes considerably due to the $\delta$ meson and its mixing, as already explained in Figure~\ref{fig:Pressure}.
Note the similar growth of $c_{s}$ is also seen in QHC21 based on the framework of quark--hadron crossover~\citep{Kojo:2021wax}.

%%%%%%%%%%%%%%%%%%%%%%%%%%%%%%%%%%%%%%%%%%%%%%%%%%%%%%%%%%%%%%%%%%%%%%%%%%%%%%%% 
\begin{figure}[t!]
  \plotone{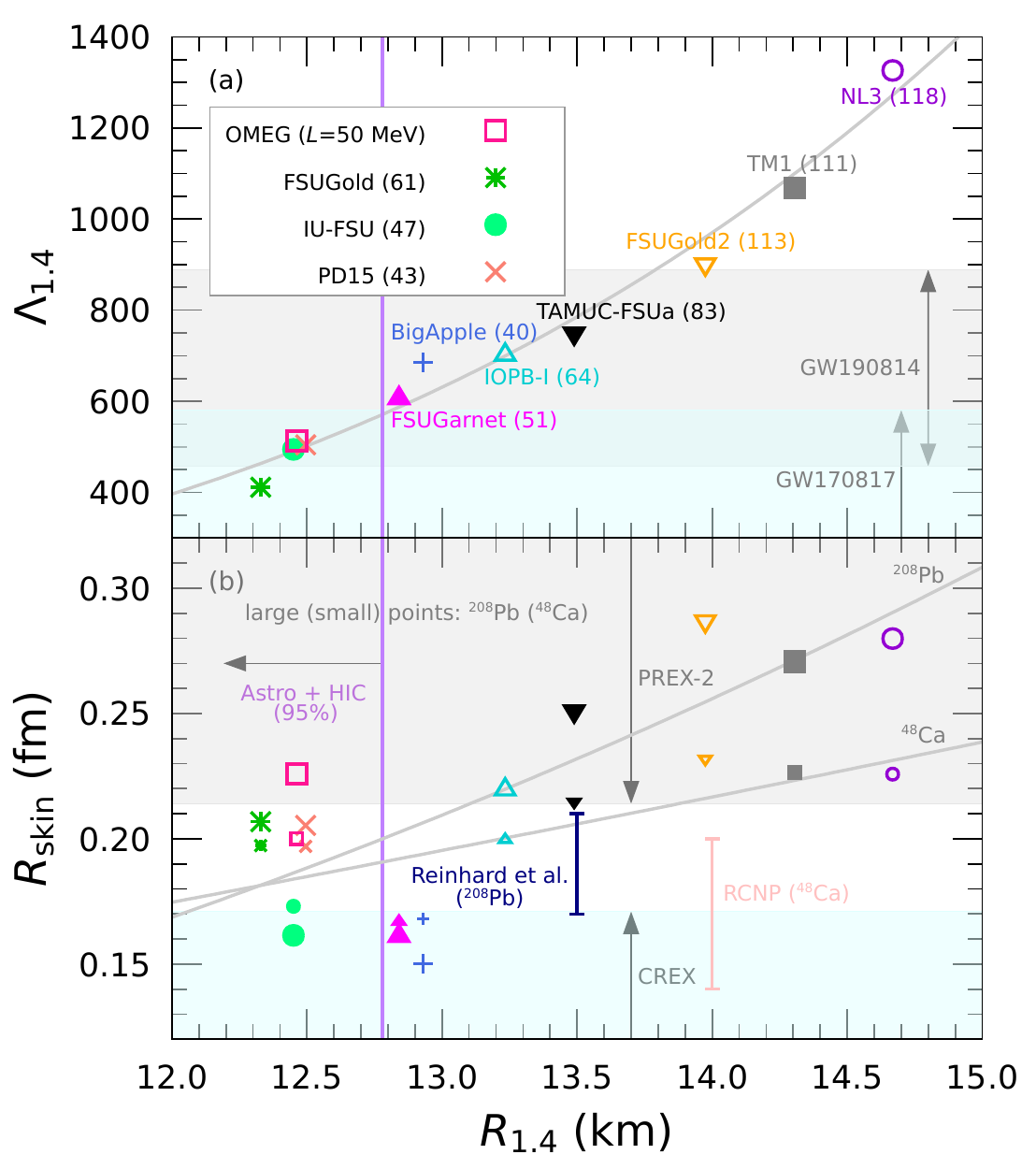}
  \caption{\label{fig:Rskin}
    (a) Correlation between $\Lambda_{1.4}$ and $R_{1.4}$.
    The solid line represents the fit $\Lambda_{1.4}=2.12\times10^{-4}$($R_{1.4}/$km)$^{5.81}$, and the values of $L$ are given in the parentheses.
    We present the constraints on $\Lambda_{1.4}$ from GW170817 ($\Lambda_{1.4}=190^{+390}_{-120}$)~\citep{LIGOScientific:2018cki} and GW190814 ($\Lambda_{1.4}=616^{+273}_{-158}$)~\citep{LIGOScientific:2020zkf}.
    (b) Correlation between $R_{\rm skin}$ and $R_{1.4}$.
    The large (small) points express the calculated values of $R_{\rm skin}$ in $^{208}$Pb ($^{48}$Ca).
    The solid lines are, respectively, the fitting functions of $R_{\rm skin}/$fm$=2.04\times10^{-4}$($R_{1.4}/$km)$^{2.70}$ for $^{208}$Pb and $R_{\rm skin}/$fm$=5.41\times10^{-4}$($R_{1.4}/$km)$^{1.40}$  for $^{48}$Ca.
    The experimental data are given as follows:
    the analytical results obtained by the PREX collaboration ($R_{\rm skin}=0.283\pm0.071$ fm)~\citep{PREX:2021umo} and \citet{Reinhard:2021utv} ($R_{\rm skin}=0.19\pm0.02$ fm) are for the case of $^{208}$Pb, while the CREX ($R_{\rm skin}=0.121\pm0.026\pm0.024$ fm)~\citep{CREX:2022kgg} and RCNP ($R_{\rm skin}=0.14$--$0.20$ fm)~\citep{Birkhan:2016qkr} results are for $^{48}$Ca.
    The purple lines in both panels correspond to the theoretical upper limit of $R_{1.4}$ given by the recent calculation using Bayesian inference with the $95\%$ credible interval~\citep{Huth:2021bsp}.}
\end{figure}
%%%%%%%%%%%%%%%%%%%%%%%%%%%%%%%%%%%%%%%%%%%%%%%%%%%%%%%%%%%%%%%%%%%%%%%%%%%%%%%% 
In the upper panel of Figure~\ref{fig:Rskin}, we present the correlation between $\Lambda_{1.4}$ and $R_{1.4}$.
As explained in \citet{Nandi:2018ami}, $\Lambda_{1.4}$ strongly interrelates with $R_{1.4}$.
In addition, $\Lambda_{1.4}$ is roughly associated with $L$ via $R_{1.4}$.
We find that the observed $\Lambda_{1.4}$ from GW170817 favors the small $R_{1.4}$ and hence $L$ too.
If both restrictions on $\Lambda_{1.4}$ from GW170817 and GW190814 are taken into account, only the IF-FSU, PD15, and OMEG interactions are acceptable as the appropriate EoS for neutron stars.
It is thus possible to mention that $R_{1.4}$ lies around 12.5 km and $L$ is in the range of $40\le L\text{(MeV)}\le50$.

The lower panel of Figure~\ref{fig:Rskin} shows the correlation between $R_{\rm skin}$ and $R_{1.4}$.
In general, the larger $R_{1.4}$ provides the thicker $R_{\rm skin}$ of $^{208}$Pb in the usual RMF models.
To describe the PREX-2 result, $L$ is thus larger than 64 MeV, given by the IOPB-I interaction.
Meanwhile, only the OMEG interaction can support the PREX-2 data with the small $L$ ($=50$ MeV) due to the $\sigma$--$\delta$ mixing.
In addition, it almost fulfills the experimental result of $R_{\rm skin}$, implied by the determination of electric dipole strength distribution in $^{48}$Ca at RCNP~\citep{Birkhan:2016qkr}.
However, it is difficult to explain the latest data reported by the CREX collaboration~\citep{CREX:2022kgg}.
From this fact, we may infer that the results of the PREX-2 and CREX experiments seem to be incompatible in the present calculations.

\section{Summary and conclusion} \label{sec:summary}

Using the RMF model with the isoscalar- and isovector-meson mixing, $\sigma^{2}\bm{\delta}^{2}$ and $\omega_{\mu}\omega^{\mu}\bm{\rho}_{\nu}\bm{\rho}^{\nu}$, we have presented an EoS for nuclear and neutron star matter, which can explain both data from terrestrial experiments and astrophysical observations of neutron stars.
A new effective interaction, dubbed the OMEG interaction, has been constructed so as to reproduce the saturation condition of nuclear matter and the ground-state properties of finite nuclei.
In particular, we have then focused on the recent experimental and observational results: the $R_{\rm skin}$ of $^{208}$Pb reported by the PREX collaboration~\citep{PREX:2021umo}, the $\Lambda_{1.4}$ observed from GW170817~\citep{LIGOScientific:2018cki}, and a $2.6$ $M_{\odot}$ compact star implied by the secondary object of GW190814~\citep{LIGOScientific:2020zkf}.

We have demonstrated that the OMEG interaction successfully accounts for the binding energies and charge radii of several closed-shell nuclei.
Especially, the thick $R_{\rm skin}$ measured by the PREX-2 experiment is achieved with small $L$ using the $\sigma$--$\delta$ mixing.
Compared with the well-calibrated RMF models, $E_{\rm sym}$ becomes very soft at high densities because the $\sigma$--$\delta$ mixing strongly affects its density dependence.
Furthermore, the neutron star EoS for the OMEG interaction is soft up to $1.5\rho_{0}$, and exhibits the rapid stiffening around $2\rho_{0}$.
It is then found that $M_{\rm max}$ exceeds the recently observed massive neutron star, PSR J0952$-$0607, which has the largest well-measured mass of $2.35\pm0.17$ $M_{\odot}$ with small modeling uncertainties~\citep{Romani:2022jhd}.
Moreover, it is possible for the OMEG interaction to satisfy the constraints on $\Lambda_{1.4}$ observed from GW170817 and GW190814.
Finally, we have thus speculated that the secondary component of GW190814 with the mass of $2.6$ $M_{\odot}$ is the heaviest neutron star ever discovered.

\vskip\baselineskip

We thank H.~Sagawa for informative discussions of the PREX-2 result.
This work is supported by the National Research Foundation of Korea (Grant Nos. NRF-2021R1A6A1A03043957, NRF-2020K1A3A7A09080134, and NRF-2020R1A2C3006177) and a NRF grant funded by the Korea government (MSIT) (No. 2018R1A5A1025563).

\bibliography{OMEG7.bib}{}
\bibliographystyle{aasjournal}

\end{document}